\documentclass[10pt, journal, final]{IEEEtran}
\usepackage{amssymb} \usepackage{cite} \usepackage{amsmath}
\interdisplaylinepenalty=2500 \usepackage{subfigure} \usepackage{graphicx}
\usepackage[dvipsnames]{xcolor} 
\usepackage{algorithm} \usepackage{algorithmic}

\newcommand{\VEC}[1]{\boldsymbol{#1}} \newcommand{\MAT}[1]{\boldsymbol{#1}}

\newcommand{\vh}{\VEC{h}} \newcommand{\vy}{\VEC{y}} \newcommand{\vn}{\VEC{n}}
\newcommand{\vg}{\MAT{G}} \newcommand{\vu}{\VEC{U}} \newcommand{\vw}{\VEC{w}}
\newcommand{\vx}{\VEC{x}}  \newcommand{\vm}{\VEC{m}}
\newcommand{\vbg}{\VEC{g}}
\newcommand{\vbu}{\VEC{u}}

\newcommand{\mH}{\MAT{H}} \newcommand{\mZ}{\MAT{Z}} \newcommand{\mK}{\MAT{K}}
 \newcommand{\mI}{\MAT{I}} \newcommand{\mQ}{\MAT{Q}}
\newcommand{\mF}{\MAT{F}} \newcommand{\mW}{\MAT{W}} \newcommand{\mB}{\MAT{B}}
\newcommand{\mA}{\MAT{A}} 
\newcommand{\mG}{\MAT{G}}

\usepackage{color,soul}

\newcommand{\ud}{\mathrm{d}}
\newcommand{\uT}{\mathrm{T}}
\newcommand{\uH}{\mathrm{H}}

\title{A Message-Passing Approach for Joint Channel
  Estimation, Interference Mitigation and Decoding}  
\author{Yan Zhu,~\IEEEmembership{Student~Member,~IEEE,}
    Dongning Guo,~\IEEEmembership{Member,~IEEE,} and Michael L. Honig,~\IEEEmembership{Fellow,~IEEE}%
\thanks{Manuscript received December 28 2008; revised July 07 2009, September 11 2009; accepted September 23 2009.  The associate editor coordinating the review of this paper and approving it for publication was Matthew Valenti.}%
\thanks{The authors are with the Department of Electrical Engineering and Computer Science, Northwestern University, Evanston, IL 60208, USA. (Email: \{yan-zhu, dguo\}@northwestern.edu, mh@eecs.northwestern.edu)}
\thanks{This work has been presented in part at
the 2008 IEEE ICC, Beijing, China and the XXIXth URSI General
Assembly, Chicago, IL, USA. }
\thanks{This work was supported in part by the Northwestern Motorola Center for Seamless Communications.}}
\begin{document}
\maketitle
\begin{abstract}
  Channel uncertainty and co-channel interference are two major challenges in
  the design of wireless systems such as future generation cellular networks.
  This paper studies receiver design for a wireless channel model with both
  time-varying Rayleigh fading and strong co-channel interference of similar
  form as the desired signal.  It is assumed that the channel coefficients of
  the desired signal can be estimated through the use of pilots, whereas no pilot
  for the interference signal is available, as is the case in many practical
  wireless systems.  Because the interference process is non-Gaussian, treating
  it as Gaussian noise generally often leads to unacceptable performance.  In
  order to exploit the statistics of the interference and correlated fading in
  time, an iterative message-passing architecture is proposed for joint channel
  estimation, interference mitigation and decoding.  Each message takes the form
  of a mixture of Gaussian densities where the number of components is limited
  so that the overall complexity of the receiver is constant per symbol
  regardless of the frame and code lengths.  Simulation of both coded and
  uncoded systems shows that the receiver performs significantly better than
  conventional receivers with linear channel estimation, and is robust with
  respect to mismatch in the assumed fading model.
\end{abstract}

\section{Introduction}\label{sec:intro}

With sufficient signal-to-noise ratio, the performance of a wireless terminal is
fundamentally limited by two major factors, namely, interference from other
terminals in the system and uncertainty about channel
variations~\cite{Tse05book}.  Although each of these two impairments has been
studied in depth assuming the absence of the other, much less is understood when
both are significant. This work considers the detection of one digital signal in
the presence of correlated fading and an interfering signal of the same
modulation type, possibly of similar strength, and also subject to independent
time-correlated fading.  Moreover, it is assumed that the channel condition of
the desired user can be measured using known pilots interleaved with data
symbols, whereas no pilot from the interferer is available at the receiver. The
practical motivation for this problem is the orthogonal frequency-division
multiple access~(OFDMA) downlink with a single dominant co-channel interferer in
an adjacent cell, which is typical in fourth generation cellular networks. Such
a situation also arises, for example, in peer-to-peer wireless networks.

This work focuses on a narrowband system with binary phase shift keying (BPSK)
modulation, where the fading channels of the desired user and the interferer are
modeled as independent Gauss-Markov processes.\footnote{The desired user and the
  interferer are modeled as independent. In principle, the fading statistics can be
  estimated and are not needed {\em a priori}.} A single transmit antenna and
multiple receive antennas are assumed first to develop the receiver, while
extensions to more elaborate models are also discussed.

The unique challenge posed by the model considered is the simultaneous
uncertainty associated with the interference and fading channels. A conventional
approach is to first measure the channel state (with or without interference),
and then mitigate the interference assuming the channel estimate is exact. Such
separation of channel estimation and detection is viable in the current problem
if {\em known} pilots are also embedded in the interference. As was shown
in~\cite{Sil07OFDM}, knowledge of pilots in the interfering signal can be
indispensable to the success of linear channel estimation, even with iterative
Turbo processing.  Without such knowledge, linear channel estimators, which
treat the interference as white Gaussian noise, provide inaccurate channel
estimates and unacceptable error probability in case of moderate to strong
interference.

Evidently, an alternative approach for joint channel estimation and interference
mitigation is needed. In the absence of interfering pilots, the key is to
exploit knowledge of the non-Gaussian statistics of the interference. The
problem is basically a compound hypothesis testing problem (averaged over
channel uncertainty).  Unfortunately, the Maximum Likelihood (ML) detector
becomes computationally impractical since it must search over (possibly a
continuum of) combined channel and interference states for all interferers.

In this paper, we develop an iterative message-passing algorithm for joint
channel estimation and interference mitigation, which can also easily
incorporate iterative decoding of error-control codes.  The algorithm is based
on belief propagation (BP), which performs statistical inference on
\emph{graphical models} by propagating locally computed ``beliefs''
\cite{Kschischang01factor}. BP has been successfully applied to the decoding of
low-density parity-check (LDPC) codes\cite{Richardson01LDPC1,
  Richardson01LDPC2}.  Other related applications of BP include combined channel
estimation and detection for a single-user fading channel or frequency selective
channel \cite{Worthen01FG, Jin08LDPC, Komninakis01Joint, Guo08LMMSE}, multiuser
detection for CDMA with ideal (nonfading) channels based on a factor graph
approach \cite{Boutros02BP} (see also \cite{Guo08MUD, Guo08MUD2}), and the
mitigation of multiplicative phase noise in addition to thermal noise
\cite{Barbieri07PhaseNoise, Colavolpe05PhaseNoise, Dauwels04PhaseNoise}.  Unique
to this paper is the consideration of fading as well as the presence of a strong
interferer.  This poses additional challenges, since the desired signal has both
phase and amplitude ambiguities, which are combined with the uncertainty
associated with the interference.

The following are the main contributions of this paper:
\begin{enumerate}
\item A factor graph is constructed to describe the model, based on which a BP
  algorithm is developed. For a finite block of channel uses, the algorithm
  performs optimal detection and estimation in two passes, one forward and one
  backward.
\item For practical implementation, the belief messages (continuous densities)
  are parametrized using a small number of variables.  The resulting
  suboptimal message-passing algorithm has constant complexity per bit (unlike
  the complexity of ML which grows exponentially with the block length).
\item Decoding of channel codes of LDPC-type is also incorporated in the
  message-passing framework.
\item As a benchmark for performance, a lower bound for the optimal uncoded error
  probability is approximated by assuming a genie-aided receiver in which
  adjacent channel coefficients are revealed.
\end{enumerate}
Numerical results are presented, which show that the message-passing algorithm
performs remarkably better than the conventional technique of linear channel
estimation followed by detection of individual symbols with or without
error-control coding. Furthermore, the relative gain is not substantially
diminished in the presence of model mismatch (\emph{i.e.}, if the Markov channel
model assumed by the receiver is inaccurate), as long as the channels
do not vary too quickly.

The remainder of this paper is organized as follows. The system model is
formulated in Section~\ref{sec:model}, and Section~\ref{sec:fgraph} develops the
message-passing algorithm. A lower bound for the error probability is studied in
Section~\ref{sec:ana}.  Section~\ref{sec:ext} discusses the extensions to
general scenarios and the computational complexity of the proposed
algorithm. Simulation results are presented in Section~\ref{sec:simu} and
Section~\ref{sec:con} concludes the paper.

 
\section{System Model}\label{sec:model}
Consider a narrow-band system with a single transmit antenna and $N_R$ receive
antennas, where the received signal at time $i$ in a frame (or block) of length
$l$ is expressed as

\begin{equation}
  \label{eq:sys}
  \vy_i = \vh_i x_i + \vh'_i x'_i + \vn_i \qquad i = 1 \ldots l
\end{equation}
where $x_i$ and $x'_i$ denote the transmitted symbols of the desired user and
interferer, respectively, $\vh_i$ and $\vh'_i$ denote the corresponding
$N_R$-dimensional vectors of channel coefficients whose covariance matrices are
$\sigma_h^2 \mI$ and $\sigma_{h'}^2 \mI$, and $\{\vn_i \}$ represents the
circularly-symmetric complex Gaussian~(CSCG) noise at the receiver with
covariance matrix $\sigma_n^2\mI$.  For simplicity, we assume BPSK modulation,
\emph{i.e.}, $x_i, x'_i $ are i.i.d. with values $\pm 1$. 

Assuming Rayleigh fading, $\{ \vh_i \}$ and $\{ \vh'_i \}$ are modeled as two
independent Gauss-Markov processes, that is, they are generated by first-order
auto-regressive relations ({\it e.g.}, \cite{Medard05PSAM}):
\begin{subequations}
  \label{eq:auto-reg}
  \begin{align}
    \vh_i&= \alpha \vh_{i-1} + \sqrt{1-\alpha^2}\,\vw_i\\
    \vh'_i&=\alpha \vh'_{i-1} + \sqrt{1-\alpha^2}\,\vw'_i
  \end{align}
\end{subequations}
where $\{\vw_i\}$ and $\{\vw'_i\}$ are independent white CSCG processes with
covariance $\sigma_h^2 \mI$ and $\sigma_{h'}^2 \mI$, respectively, and $\alpha$
determines the correlation between successive fading coefficients.  This model
includes two special cases: $\alpha=0$ corresponds to independent fading and
$\alpha=1$ corresponds to block fading. Although this model is simple, general
fading model can be approximated by such first-order Markovian
model~\cite{Wang96MC, Tan00Channel} via choosing appropriate value for
$\alpha$. Furthermore, numerical simulations in Section~\ref{sec:simu} also show
that the receiver designed under such channel assumption is robust in other
fading environments as long as the channel variation over time is not too
fast. Note that \eqref{eq:sys} also models an OFDM system where $i$ denotes the
index of sub-carriers instead of the time index.

Typically, pilots are inserted periodically among data symbols. For example,
25\% pilots refers to pattern ``PDDDPDDDPDDD...'', where P and D mark pilot and
data symbols, respectively.  Let $\vy_i^j$ denote the sequence $\vy_i, \vy_{i+1},
\ldots, \vy_j$.  The detection problem can be formulated as follows: Given the
observations $\vy_1^l$ and known value of a certain subset of symbols in $x_1^l$
which are pilots, we wish to recover the information symbols from the desired
user, \emph{i.e.}, the remaining unknown symbols in $x_1^l$, where the realization of
the channel coefficients and interfering symbols is not available.
  
\section{Graphical Model and Message Passing}\label{sec:fgraph}
\subsection{Graphical Model for Uncoded System}
\begin{figure}[h]
  \centering
  \includegraphics[width=3.5in]{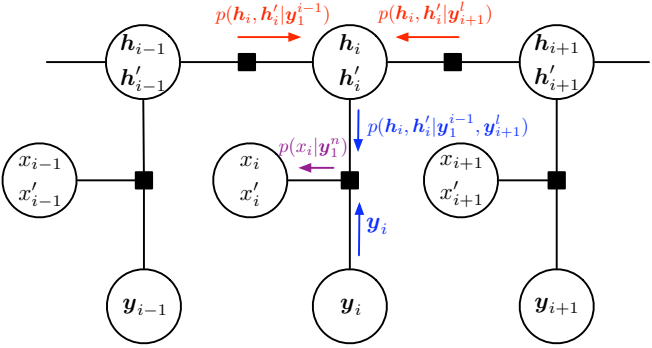}
  \caption{A factor graph describing the communication system model without
    channel coding. The arrows refer to messages which are discussed in
    Section~\ref{sec:fgraph_ex}}
  \label{fig:hmc}
\end{figure}

An important observation from \eqref{eq:sys} and~\eqref{eq:auto-reg} is that the
fading coefficients $\{(\vh_i,\vh_i')\}_{i=1}^l$ form a Markov chain with state
space in $\mathbb{C}^{2N_r}$. Also, given $\{(\vh_i,\vh'_i)\}_{i=1}^l$, the
  3-tuple $(x_i, x'_i, \vy_i)$ of input and output variables is independent
over time $i=1,2,\ldots,l$.  The joint distribution of the random variables can
be factored as
\begin{multline*}
  p(\vy_1^l, x_1^l, x_1'^l, \vh_1^l, \vh_1'^l) = p(\vh_1,\vh'_1)\prod_{i=2}^l
  p(\vh_i,\vh'_i|\vh_{i-1}, \vh'_{i-1}) \\ \times \prod_{i=1}^l
  \Big( p(\vy_i|\vh_i,\vh'_i,x_i,x_i')p(x_i)p(x_i') \Big)\,.
\end{multline*}
This factorization can be described using the {\em factor graph} shown in Fig.~\ref{fig:hmc}.

Generally, a factor graph is a \emph{bipartite graph}, which consists of two
types of nodes: the variable nodes, each denoted by a circle in the graph, which
represents one or a few random variables jointly; and the factor nodes, each
denoted by a square which represents a constraint on the variable nodes
connected to it~\cite{Kschischang01factor, Loeliger07FGraph}. The factor node
between the node $(\vh_i,\vh'_i)$ and the node $(\vh_{i-1},\vh'_{i-1})$
represents the conditional distribution $p(\vh_i,\vh'_i|\vh_{i-1}, \vh'_{i-1})$,
which is the probability constraint specified by~\eqref{eq:auto-reg}. Similarly,
the factor node connecting nodes $\vy_i$, $(\vh_i,\vh'_i)$ and $(x_i,x_i')$
represents the conditional distribution $p(\vy_i|\vh_i,\vh'_i,x_i,x_i')$, which
is the relation given by~\eqref{eq:sys}.  The prior probability distribution of
the data symbols is assigned as follows.  All BPSK symbols $x_i$ and $x'_i$ are
uniformly distributed on $\{-1,1\}$ except for the subset of pilot symbols in
$x^l_1$, which are set to 1. The
Markovian property of the graph is that conditioned on any cut node(s), the
separated subsets of variables are mutually independent. As we shall see, the
Markovian property plays an important role in the development of the
message-passing algorithm.

Since the graphical model in Fig.~\ref{fig:hmc} fully describes the probability
laws of the random variables given in (\ref{eq:sys}) and (\ref{eq:auto-reg}),
the detection problem is equivalent to statistical inference on the
graph\footnote{The style of the factor graph in Fig.~\ref{fig:hmc} is similar
  to that used in some other work, such as~\cite{Jin08LDPC},
  \cite{Barbieri07PhaseNoise} and~\cite{Niu05BP}, which address channel and
  phase uncertainty in the absence of interference. Note that there are several different styles of factor graphs,
  {\it e.g.}, the Forney style~\cite{Forney01NormGraph} which uses edges rather than
  circular nodes to represent variables.}.  Simply put, we seek to answer the
following question: Once the realization of a subset of the variables (received
signal and pilots) on the graph is revealed, what can be inferred about the
symbols from the desired user?

Note that the same factor graph would arise if we were to jointly detect both
$x_i$ and $x'_i$, which solves a problem of multiuser detection. In this work,
however, the receiver is only interested in detecting $x_i$, so that $x_i'$ is
being averaged out when passing messages between nodes $(\vh_i, \vh_i')$ and
$(\vh_{i+1}, \vh_{i+1}')$.

\subsection{Exact Inference via Message Passing}\label{sec:fgraph_ex}
In the problem described in Section \ref{sec:model}, the goal of inference is to
obtain or approximate the marginal posterior probability $p(x_i|\vy_1^l)$, which
is in fact a sufficient statistic of $\vy_1^l$ for $x_i$. Problems of such
nature have been widely studied (see e.g., \cite[Chapter 4]{MacKay03book} and
\cite{Kschischang01factor}).  In particular, BP is an efficient algorithm for
computing the posteriors by passing messages among neighboring nodes on the
graph. In principle, the result of message passing with sufficient number of
steps gives the exact {\em a posteriori} probability of each unknown random
variable if the factor graph is a tree ({\em i.e.}, free of cycles). For general
graphs with few short cycles, iterative message-passing algorithms often compute
good approximations of the desired probabilities.  Unlike in most other work
(including~\cite{Barbieri07PhaseNoise,Colavolpe05PhaseNoise,Dauwels04PhaseNoise}),
where each random variable is made a variable node, multiple variables are
clustered into a single node so that the factor graph in Fig.~\ref{fig:hmc} is
free of cycles (see also~\cite{Kschischang01factor} for the usage of the
clustering techniques). Numerical experiments (omitted here) show that making
each variable a separate node leads to poor performance due to a large number of
short cycles ({\it e.g.}, there will be a cycle through $\vh_i, \vh_i,
\vh_{i+1}, \vh_{i+1}'$).

Let $\vg_i=[\vh_i, \vh'_i]$ and $\vu_i = [\vw_i, \vw'_i]$.  The model
(\ref{eq:sys}) and (\ref{eq:auto-reg}) can be rewritten as:
\begin{align}
  \vy_i &= \vg_i
  \begin{bmatrix}
    x_i \\
    x'_i
  \end{bmatrix}
+ \vn_i \label{eq:sys2}\\
  \vg_{i} &= \alpha \vg_{i-1} + \sqrt{1-\alpha^2}\, \vu_i. \label{eq:auto-reg2}
\end{align}

The probability distributions immediately available are $p(\vy_i| \vg_i, x_i,
x'_i )$, $p(\vg_i| \vg_{i-1})$ and the marginals $p(x_i)$, $p(x'_i)$, as well as
$p(\vg_i)$ which are Gaussian. Note that $p(\vy_i|\vg_i,x_i, x'_i)$ is the
conditional Gaussian density corresponding to the channel model (\ref{eq:sys})
and $p(x_i, x'_i)=p(x_i)p(x'_i)$ since the desired symbol and the interference
symbol are independent. Also, $P(x_i=1)=1$ and $P(x_i=-1)=0$ if $x_i$ is a pilot
for the desired user, otherwise $P(x_i=\pm 1)=1/2$.  Moreover,
$P(x'_i=\pm1)\equiv 1/2$ for all $i$, since we do not know the pilot pattern of
the interfering user.

The goal is to compute for each $i=1,\dots,l$:
\begin{align*}
  p(x_i|\vy_1^l) &= \sum_{x'_i=\pm 1} \int p(x_i, x'_i, \vg_i|\vy_1^l) \,\ud
  \vg_i \\
  &\propto \sum_{x'_i=\pm 1} \int p(x_i, x'_i, \vy_1^{i-1}, \vy_i, \vy_{i+1}^l, \vg_i) \, \ud \vg_i
\end{align*}
where the ``proportion'' notation $\propto$ indicates that the two sides differ
by a factor which depends only on the observation $\vy^l_1$ (which has no
influence on the likelihood ratio $ P(x_i=1|\vy_1^l) / P(x_i=-1|\vy_1^l) $ and
hence on the decision). For notational simplicity we have also omitted the
limits of the integrals, which are over the entire axes of $2N_R$ complex
dimensions. By the Markovian property, $(x_i, x'_i, \vy_i)$, $\vy_1^{i-1}$ and
$\vy_{i+1}^l$ are mutually independent given $\vg_i$.  Therefore,
\begin{align}
  p(x_i|\vy_1^l) &\propto \sum_{x'_i=\pm 1} \int p(\vy_i, x_i, x'_i | \vg_i)
  p(\vy_1^{i-1}|\vg_i) \nonumber \\ 
  &\hspace{10em} \times p(\vy_{i+1}^l|\vg_i)
  p(\vg_i)\, \ud \vg_i \nonumber \\
  &\propto \! \sum_{x'_i=\pm 1} \! p(x_i) p(x'_i) \!\!\!\int
  \!\!p(\vy_i|\vg_i,x_i, x'_i) p(\vg_i|\vy_1^{i-1}) \nonumber \\
  &\hspace{8em} \times p(\vg_i|\vy_{i+1}^l)\big/
  p(\vg_i) \ud \vg_i\label{eq:soft}
\end{align}
where the independence of $(x_i, x'_i)$ and $\vg_i$ is used to
obtain~\eqref{eq:soft}.  In order to compute (\ref{eq:soft}), it suffices to
compute $p(\vg_i|\vy_1^{i-1})$ and $p(\vg_i|\vy_{i+1}^l)$.

We briefly derive the posterior probability $p(\vg_i|\vy^{i-1}_1)$ as a
recursion in below, whereas computation of $p(\vg_i|\vy_{i+1}^l)$ is similar by
symmetry.  Consider the posterior of the coefficients $\vg_i$ given the received
signal up to time $i-1$. The influence of $\vy_1^{i-1}$ on $\vg_i$ is through
$\vg_{i-1}$ because $\vg_i$ and $\vy_1^{i-1}$ are independent given
$\vg_{i-1}$. Thus,
\begin{equation*}
  p(\vg_i|\vy^{i-1}_1) = \int p(\vg_i|\vg_{i-1}) p(\vg_{i-1}|\vy^{i-1}_1) \ud \vg_{i-1}\,.   
\end{equation*}
By the Markovian property, $\vy^{i-2}_1$ and $\vy_{i-1}$ are independent given
$\vg_{i-1}$, so that
\begin{equation}
  p(\vg_i|\vy^{i-1}_1) \! \! \propto \!\!\! \int \!\! p(\vg_i|\vg_{i-1}) p(\vy_{i-1}|\vg_{i-1})
  p(\vg_{i-1}|\vy^{i-2}_1) \ud \vg_{i-1}. \label{eq:A}
\end{equation}
Since $\vg_{i-1}$ and $x_{i-1}, x'_{i-1}$ are independent,
\begin{align}
  p&(\vy_{i-1}|\vg_{i-1}) \nonumber \\
  &= \hspace{-1em} \sum_{x_{i-1}, x'_{i-1}=\pm 1} \hspace{-1em} p(\vy_{i-1}|x_{i-1},
  x'_{i-1},\vg_{i-1})p(x_{i-1})p(x'_{i-1})\,. \label{eq:B}
\end{align}
Therefore, by~\eqref{eq:A} and~\eqref{eq:B}, we have a recursion for computing $p(\vg_i|\vy^{i-1}_1)$ for
each $i=1,\dots,l$, 
\begin{multline}
  \label{eq:fwd_h}
  p(\vg_i|\vy^{i-1}_1) \propto \hspace{-1em} \sum_{x_{i-1}, x'_{i-1}=\pm 1} \int p(\vg_i|\vg_{i-1})
  p(\vg_{i-1}|\vy^{i-2}_1) \\
\times p(\vy_{i-1}|\vg_{i-1},x_{i-1}, x'_{i-1}) p(x_{i-1})p(x'_{i-1})\,\ud
  \vg_{i-1}
\end{multline}
which is the key to the message-passing algorithm. Similarly, we can also
derive the inference on $\vg_i$, which serves as an estimate of channel
coefficients at time~$i$:
\begin{multline}
  \label{eq:est}
  p(\vg_i|\vy_1^l) \propto \sum_{x_i, x'_i = \pm 1} p(x_i)p(x'_i)
  p(\vy_i|\vg_i,x_i, x'_i) \\
  \times p(\vg_i|\vy_1^{i-1}) p(\vg_i|\vy_{i+1}^l)/ p(\vg_i).
\end{multline}
In other words, the BP algorithm requires backward and forward message-passing
only once in each direction, which is similar to the BCJR algorithm
\cite{BCJR74}. The key difference between our algorithm and the BCJR is that the
Markov chain here has a continuous state space.\footnote{Another way to derive
  the message passing algorithm is based on the factor graph, in which the joint
  probability is factored first and then marginalized to get the associated
  posterior probability~\cite{Kschischang01factor}.}

The joint channel estimation and interference mitigation algorithm is summarized
in Algorithm~\ref{alg:bp}.  Basically, the message from a factor node to a
variable node is a summary of the extrinsic information\footnote{It is obtain by removing the posterior
probability of the variable node itself in the  {\em a posterior} probability~(APP).}~(EI) about the random
variable(s) represented by the variable node based on all observations connected
directly or indirectly to the factor node~\cite{Kschischang01factor}. For example, the message received by
node ($\vh_i,\vh'_i$) from the factor node on its left summarizes all
information about ($\vh_i,\vh'_i$) based on the observations $\vy_1, \ldots,
\vy_{i-1}$, which is proportional to $p(\vh_i,\vh'_i|\vy^{i-1}_1)$. The message
from a variable node to a factor node is a summary of the EI about the
variable node based on the observations connected to it. For example, the
message passed by node ($\vh_i,\vh'_i$) to the factor node on its left is
the EI about ($\vh_i,\vh'_i$) based on the observations $\vy_1, \ldots,
\vy_i$, {\em i.e.}, $p(\vh_i,\vh'_i|\vy^i_1)$.

\begin{algorithm}
  \caption{Pseudo code for the message-passing algorithm}
  \label{alg:bp}
  \begin{algorithmic}
    \STATE Initialization: $P(x'_i=1)=P(x'_i=-1)=1/2$ for all $i$.  The same
    probabilities are also assigned to $p(x_i)$ for all $i$ except for the pilots, for which $P(x_i=1)=1$.  For all $i$, $p(\vg_i)$ is zero mean Gaussian with variance $\mQ$.\\
    \FOR{$i = 1$ to $l$} 
    \STATE Compute $p(\vg_i|\vy^{i-1}_1)$ from (\ref{eq:fwd_h})\\
    \STATE Compute  $p(\vg_i|\vy_{i+1}^l)$ similarly to \eqref{eq:fwd_h}\\
    \ENDFOR\\
    \FOR{$i = 1$ to $l$}
    \STATE Compute $p(x_i|\vy_1^l)$
    from (\ref{eq:soft}) \\
    \ENDFOR
  \end{algorithmic}
\end{algorithm}

Given the factor graph, it becomes straightforward to write out the message
passing algorithm, which is equivalent to the sum-product
algorithm~\cite{Kschischang01factor}. Because of the simple Markovian structure
of the factor graph, we derive the algorithm using basic probability arguments
in this section.  The preceding treatment is self-contained, and the
technique also applies to other similar problems.

\subsection{Practical Issues}
Algorithm \ref{alg:bp} cannot be implemented directly using a digital computer
because the messages are continuous probability density functions (PDFs). Here
we choose to parametrize the PDFs, as opposed to quantizing the
multi-dimensional PDFs directly, which requires a large number of quantization
bins and thus high computational complexity. Also, parametrization can
characterize the PDFs exactly without introducing extra quantization error. Thus
it can achieve better performance with less complexity.  (Of course, for
hardware implementation the PDF parameters must be quantized.)

For notational convenience, we use $\vbg$ to denote the column vector formed by
stacking the columns of the matrix $\vg$, \emph{i.e.}, if $\vg_i=[\vh_i,
\vh'_i]$ is $N_R \times 2$ as defined previously then $\vbg_i=\left[ \vh_i^\uT,
\vh'^\uT_i \right]^\uT$ is $2N_R\times 1$.  We define
\begin{align*}
  \mZ_i= \left[ x_i, x'_i \right] \otimes \mI_2 = 
   \begin{bmatrix} 
    x_i &  0  & x'_i &  0\\
      0  &x_i & 0     & x'_i
    \end{bmatrix}
\end{align*}
where $\otimes$ denotes the Kronecker product and $\mI_r$ denotes the $r\times
r$ identity matrix. Then~\eqref{eq:sys2} and~\eqref{eq:auto-reg2} are equivalent
to
\begin{align}
  \vy_i &= \mZ_i \vbg_i + \vn_i \label{eq:sys3} \\
  \vbg_i & = \alpha \vbg_{i-1} + \sqrt{1-\alpha^2}\, \vbu_i \label{eq:auto-reg3} 
\end{align} 
where $\vbu_i$ is a column vector consisting of $2N_R$ independent CSCG
variables with variance $\sigma_h^2$ or $\sigma_{h'}^2$.  Let the
$r-$dimensional complex Gaussian density be denoted by
\begin{equation*}
  \mathcal{CN}(\vx,\vm,\mK) \equiv \frac{1}{\pi^r \det(\mK)} \exp \left[ -(\vx-\vm)^\uH \mK^{-1} (\vx-\vm) \right]
\end{equation*}
where $\vx_{r \times 1}$ is a column vector of complex dimension $r$, and
$\vm_{r \times 1}$ and $\mK_{r \times r}$ denote the mean and covariance matrix,
respectively. Let $\mQ=\text{diag}(\sigma^2_h, \sigma^2_h, \sigma^2_{h'}, \sigma^2_{h'})$. 
We can then write $p(\vbg_i|\vbg_{i-1})=\mathcal{CN}
(\vbg_i, \alpha \vbg_{i-1}, \sqrt{1-\alpha^2}\,\mQ)$,
$p(\vbg_i)=\mathcal{CN}(\vbg_i, 0, \mQ)$ and
$p(\vy_i|\vbg_i, x_i, x'_i) = \mathcal{CN}(\vy_i, \mZ_i\vbg_i, \sigma^2_n \mI)$.

The density functions, $p(\vbg_i|\vy_{i+1}^l)$ and $p(\vbg_i|\vy_1^{i-1})$ are
Gaussian mixtures. Note that the random variables in Fig.~\ref{fig:hmc} are
either Gaussian or discrete. The forward recursion \eqref{eq:fwd_h} for
$p(\vbg_i|\vy_1^{i-1})$ starts with a Gaussian density function. As the message
is passed from node to node, it becomes a mixture of more and more Gaussian
densities. Each Gaussian mixture is completely characterized by the amplitudes,
means and variances of its components. Therefore, we can compute and pass these
parameters instead of PDFs.

Without loss of generality, we assume that $p(\vbg_{i-1}|\vy^{i-2}_1) = \sum_j
\rho_j\,\mathcal{CN}(\vbg_{i-1}, \vm^j_{i-1}, \mK^j_{i-1})$, where non-negative
numbers $\{\rho_j\}$ satisfy $\sum_j \rho_j = 1$. Substituting into
(\ref{eq:fwd_h}), after some manipulations, we have
\begin{equation}
  \label{eq:fwd_h2}
  p(\vbg_i|\vy^{i-1}_1) \propto \sum_{j, x_{i-1}, x'_{i-1}} \hspace{-1em}\rho_j \, p(x_{i-1})p(x'_{i-1}) L(j,i)C(j,i)
\end{equation}
where
\begin{equation}
  \label{eq:like}
  \hspace{-.7em} L(j,i) = \mathcal{CN}(\mZ_{i-1} \vm^j_{i-1}, \vy_{i-1}, \sigma^2_n\mI + \mZ_{i-1} \mK^j_{i-1} \mZ_{i-1})
\end{equation}
and
\begin{equation}
  \label{eq:comp}
  C(j,i) = \mathcal{CN}\left( \vbg_i, \vm^{j,i}_i, \mK^{j,i}_i \right) 
\end{equation}
where
\begin{subequations}
  \begin{align}
    \vm^{j,i}_i &= \alpha \vm^j_{i-1} + \alpha \mK^j_{i-1} \mZ_{i-1}^\uH (
      \sigma^2_n\mI + \mZ_{i-1} \mK^j_{i-1}
      \mZ_{i-1})^{-1} \nonumber \\
    &\hspace{10em} \times ( \vy_i - \mZ_{i-1} \vm^j_{i-1} )  \label{eq:mean}\\
    \mK^{j,i}_i &= \alpha^2 \mK^j_{i-1} + \sqrt{1-\alpha^2}\,\mQ - (\alpha
    \mK^j_{i-1}\mZ_{i-1}^\uH) \nonumber \\
    &\hspace{1em} \times ( \sigma^2_n\mI + \mZ_{i-1} \mK^j_{i-1}
    \mZ_{i-1}^\uH )^{-1} (\alpha \mZ_{i-1} \mK^j_{i-1}). \label{eq:var}
  \end{align}
\end{subequations}
Basically, (\ref{eq:fwd_h2}), (\ref{eq:like}) and (\ref{eq:comp}) give an
explicit recursive computation for the amplitude, mean and variance of each
Gaussian component in message $p(\vbg_i|\vy^{i-1}_1)$. Similar computations can
be applied to $p(\vbg_i|\vy_{i+1}^l)$.

Examining (\ref{eq:mean}) and (\ref{eq:var}) more closely, ignoring the
superscripts, they are the one-step prediction equation and Riccati equations,
respectively, for the linear system defined by~\eqref{eq:sys2} and~\eqref{eq:auto-reg2} with known $\mZ_{i-1}$ \cite[Ch.3]{Anderson00book},
\cite[Sec.  IV.C]{Kschischang01factor}. Therefore, passing messages from one end
to the other can be viewed as a series of Kalman filters with different weights:
In each step, each filter performs the traditional Kalman filter for each
hypothesis of $\mZ_{i-1}$ and the filtered result is weighted by the product of
the previous weight, the posterior probability of the hypothesis,
and $L(j,i)$.\footnote{The value of 
$L(j,i)$ is given by (\ref{eq:like}) and is related to
  the difference between the filtered result and the new observation.
Prior work in which a {\em single} Kalman filter
is used for channel estimation in the absence of interference
is presented in~\cite{Heo00Iter, Niu05BP}.}

The number of Gaussian components increases exponentially in the recursive
formula~(\ref{eq:fwd_h2}), which becomes computationally infeasible. In this
work, we fix the total number of components and simply pick the components with
the largest amplitudes (which correspond to the most likely hypotheses). In
general, this problem is equivalent to the problem of survivor-reduction. Two
techniques that have been proposed are decision feedback~\cite{Simmons90Trellis}
and thresholding~\cite{Dai94Detection}. The former limits the maximum number of
survivors by assuming the past decisions are correct, while the latter keeps the
survivors only when their {\em a posteriori} probabilities exceed a certain
threshold value. According to the preceding analysis, the method we propose
falls into the decision feedback category. Obviously, the more components we
keep, the better performance we have; however, the higher the complexity at the
receiver. We investigate this issue numerically in Section~\ref{sec:simu}. A
different approach to limiting the number of Gaussian components is presented in
\cite{Dauwels06ParticleFilter, Scott01Mixture,
  Sudderth03BP,Colavolpe05PhaseNoise ,Kurkoski08BP}.  There the basic idea is to
merge components ``close'' to each other instead of discarding the weakest ones
as we do here.  However, that requires computing distances between pairs of
components, which can lead to significantly higher complexity
\cite{Kurkoski08BP,Scott01Mixture}.  The relative performance of these different
methods is left for future study.

\subsection{Integration with Channel Coding}

\begin{figure*}
  \centering
  \includegraphics[width=5.5in]{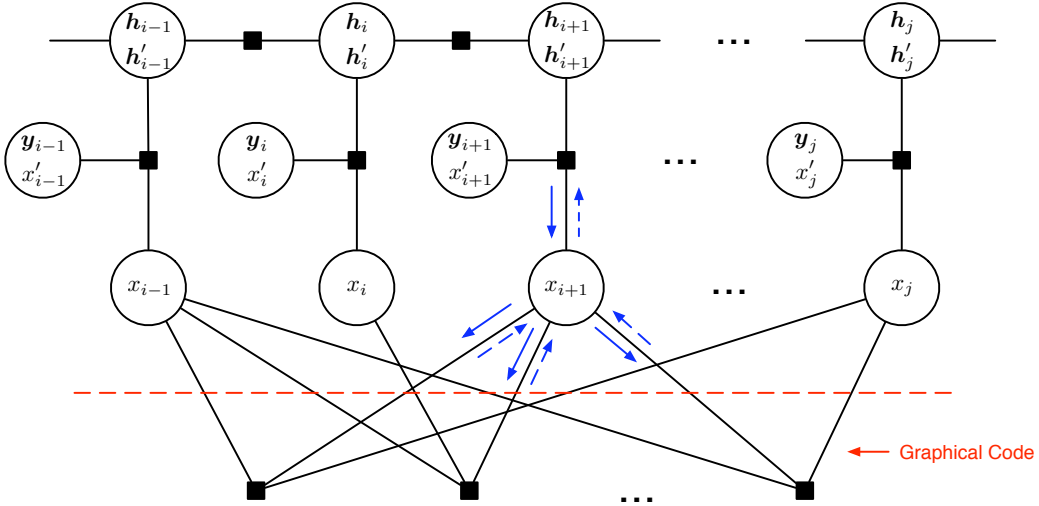}
  \caption{A factor graph for joint detection, estimation and decoding.  The
    solid arrows show EI passed from the detector to the decoder, and the dashed
    arrows show EI passed from the decoder to the detector.}
  \label{fig:bp_ldpc}
\end{figure*}

Channel codes based on factor graphs can be easily incorporated in the
message-passing framework developed thus far.  In Fig.~\ref{fig:bp_ldpc}, a
sparse graphical code is in conjunction with the factor graph for the model
(\ref{eq:sys}) and (\ref{eq:auto-reg}).  The larger factor graph is no longer
acyclic. Therefore, the message-passing algorithm is sub-optimal for this graph
even if one could keep all detection hypotheses ({\em i.e.}, the number of
mixture components is unrestrained). However, design of channel coding can
guarantee the degrees of such cycles are typically quite large.  Hence
message-passing performs very well~\cite{Richardson01LDPC2}. Based on the factor
graph, one can develop many message-passing schedules. To exploit the slow
variation of the fading channel, the non-Gaussian property of the interfering
signal and the structure of graphical codes, a simple idea is to allow the
detector and decoder to exchange their extrinsic information. For example,
suppose that at a certain message-passing stage, the node $x_{i+1}$ computes its
APP from the detector. Then the node $x_{i+1}$ can distribute the EI to the
sub-graph of the graphical code, which is described by the solid arrows in
Fig.~\ref{fig:bp_ldpc}. After $x_{i+1}$ collects the ``beliefs'' from all its
edges, it passes the EI (which is obtained by multiplying together all
``beliefs'' but the one coming from the detector) back to the detector. This
process is described by the dashed arrows in Fig.~\ref{fig:bp_ldpc}. In other
words, both the detector and the decoder compute their posterior probabilities
from received EI.

In this paper, we use LDPC codes with the following simple strategy: We run the
detection part as before and then feed the EI to the LDPC decoder through
variable nodes $x_i$. After running the LDPC decoder several rounds, we feed
back the EI to the detection sub-graph.  We investigate the impact of the
message-passing schedule on performance numerically in Section~\ref{sec:simu}.

 
\section{Error Floor Due to Channel Uncertainty}\label{sec:ana} 
Channel variations impose a fundamental limit on the error performance
regardless of the signal-to-noise ratio~(SNR). Consider a genie-aided receiver:
when detecting symbol $x_i$, a genie reveals all channel coefficients but
$(\vh_{i}, \vh'_{i})$ to the receiver, which can only reduce the minimum error
probability. Even in the absence of noise, the receiver cannot estimate $(\vh_{i},
\vh'_{i})$ precisely (not even the sign) due to the Markovian property
in~\eqref{eq:auto-reg}. Therefore, the error probability does not vanish 
as the noise power goes to zero.

Evidently, the genie-aided receiver also gives a lower bound on the error
probability for the exact message passing algorithm. In the following, we derive
an approximation to this lower bound. Numerical results in Section V indicate
that the difference between the approximate lower bound and the actual
genie-aided performance is small.

Consider the error probability of jointly detecting $[x_i,x'_i]$ with the help
of the genie.  Conditioned on all other channel coefficients, $\vh_i$ and
$\vh'_i$ are Gaussian. Let $\vh_i = \hat{\vh}_i + \tilde{\vh}_i$ and $\vh'_i =
\hat{\vh}'_i + \tilde{\vh}'_i$ where $\hat{\vh}_i$ and $\hat{\vh}'_i$ are the
estimates of $\vh_i$ and $\vh'_i$, respectively, and $\tilde{\vh}_i$ and
$\tilde{\vh}_i$ are the respective estimation errors. By treating
$\tilde{\vh}_ix_i$ and $\tilde{\vh}'_ix'_i$ as additional noise, the channel
model can be rewritten as
\begin{align*}
  \vy_i = \hat{\vh}_ix_i + \hat{\vh}'_ix'_i + \tilde{\vn}_i
\end{align*}
where the residual noise $\tilde{\vn}_i=\tilde{\vh}_ix_i + \tilde{\vh}'_ix'_i+
\vn_i$. It can be shown that $\tilde{\vn}_i$ is a CSCG random vector independent
of $(x_i, x'_i)$ and with covariance matrix $\sigma_{\tilde{n}}^2 \mI =\left(
  \frac{1 - \alpha^2}{ 1 + \alpha^2}( \sigma_h^2 + \sigma_{h'}^2 ) + \sigma_n^2
\right) \mI$.

Let $\hat{x}_i$ and $\hat{x}'_i$ be the estimates of $x_i$ and $x'_i$,
respectively. For simplicity, we assume dual receive antennas~($N_R=2$). Following a standard analysis~\cite[App. A]{Tse05book}, we have
\begin{align}  \label{eq:error} \nonumber
  \hspace{-.5em}P(\text{error}) & = P(\hat{x}_i\neq x_i, \hat{x}'_i=x'_i)
 + P( \hat{x}_i\neq x_i, \hat{x}'_i\neq x'_i)\\
           & =  \left( \frac{1-\mu_1}{2} \right)^2 \!\!(2+\mu_1) + \left(
             \frac{1-\mu_2}{2} \right)^2 (2+\mu_2)
\end{align}
where
\begin{align*}
  \mu_1 &= \sqrt{ \frac{\alpha^2
      \sigma_h^2}{\alpha^2\sigma_h^2+(1+\alpha^2)\sigma_{\tilde{n}}^2} }\\
  \mu_2 &= \sqrt{ \frac{\alpha^2 (\sigma_h^2 + \sigma_{h'}^2) }{\alpha^2
      (\sigma_h^2 + \sigma_{h'}^2) + (1+\alpha^2) \sigma_{\tilde{n}}^2} } \,.
\end{align*}
Note that as long as $\alpha \neq 1$, the residual noise $\tilde{n}$ does not
vanish, which results in an error floor. Therefore, such error floor is inherent
to the channel model, and despite its simplicity, the channel cannot be tracked
exactly based on pilots.

\begin{figure*}[t]
  \centering
    \subfigure[]{
      \label{fig:ber_w}
      \includegraphics[width=2.3in]{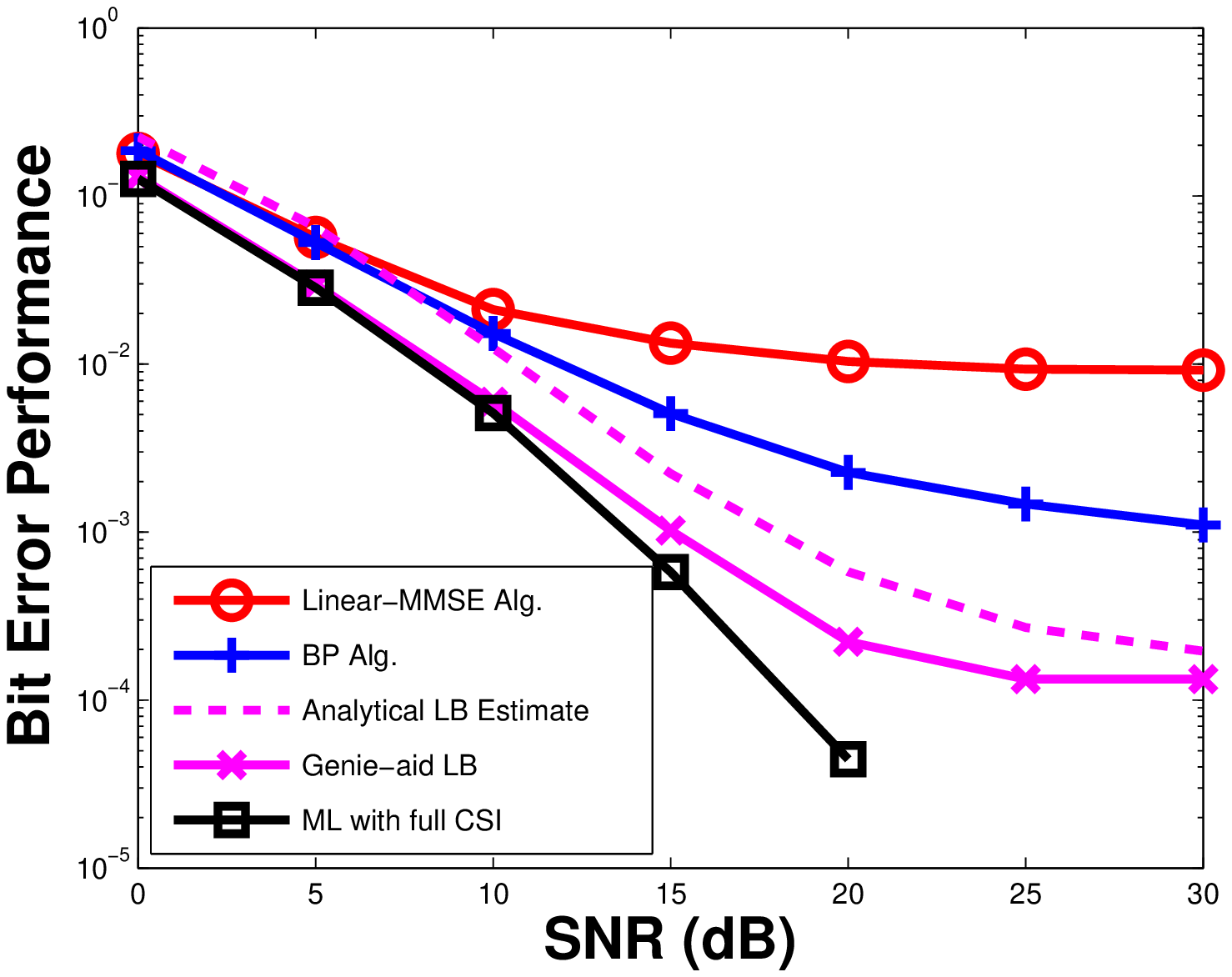}} 
    \subfigure[]{
      \label{fig:ber_m}
      \includegraphics[width=2.3in]{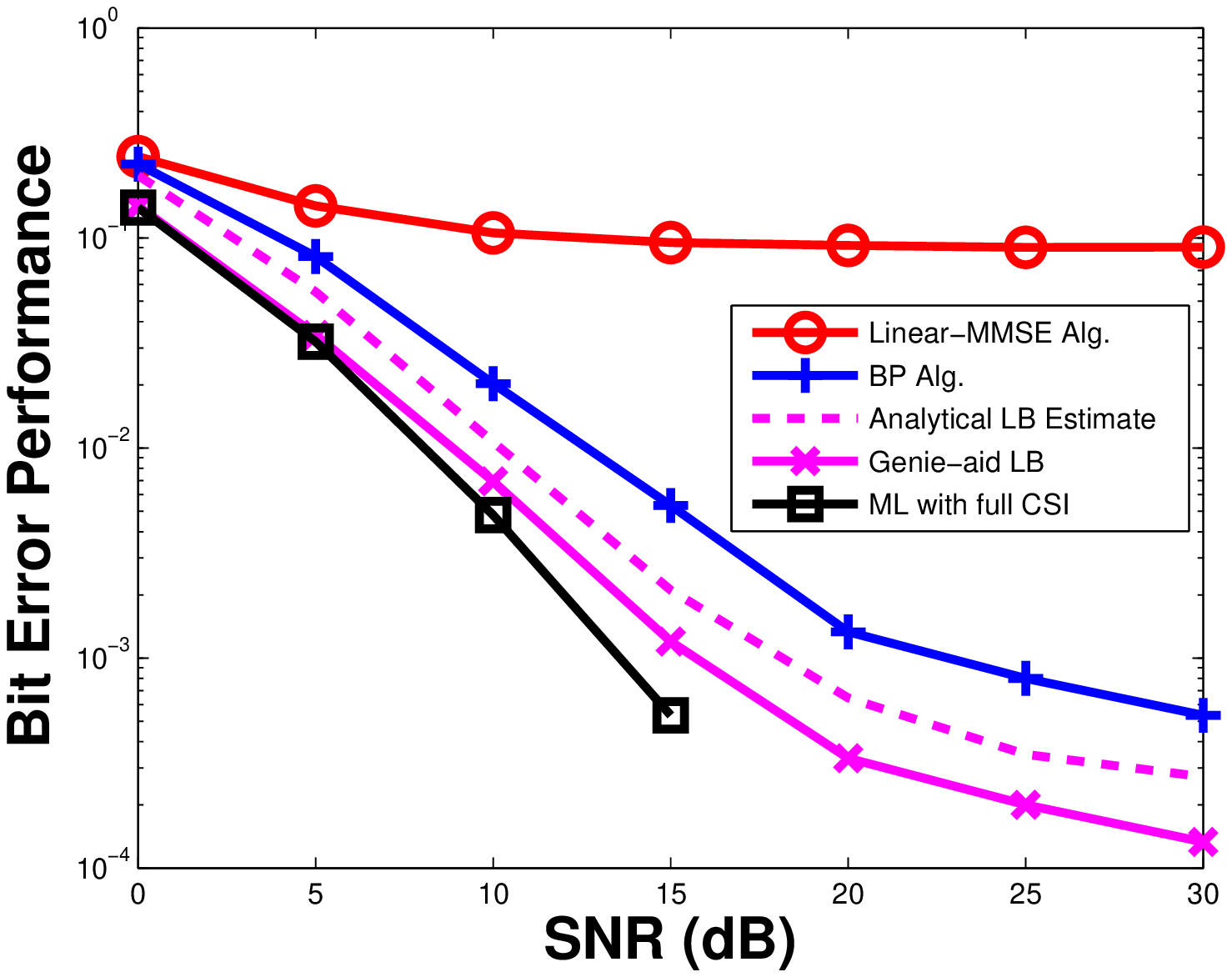}} 
    \subfigure[]{
      \label{fig:ber_s}
      \includegraphics[width=2.3in]{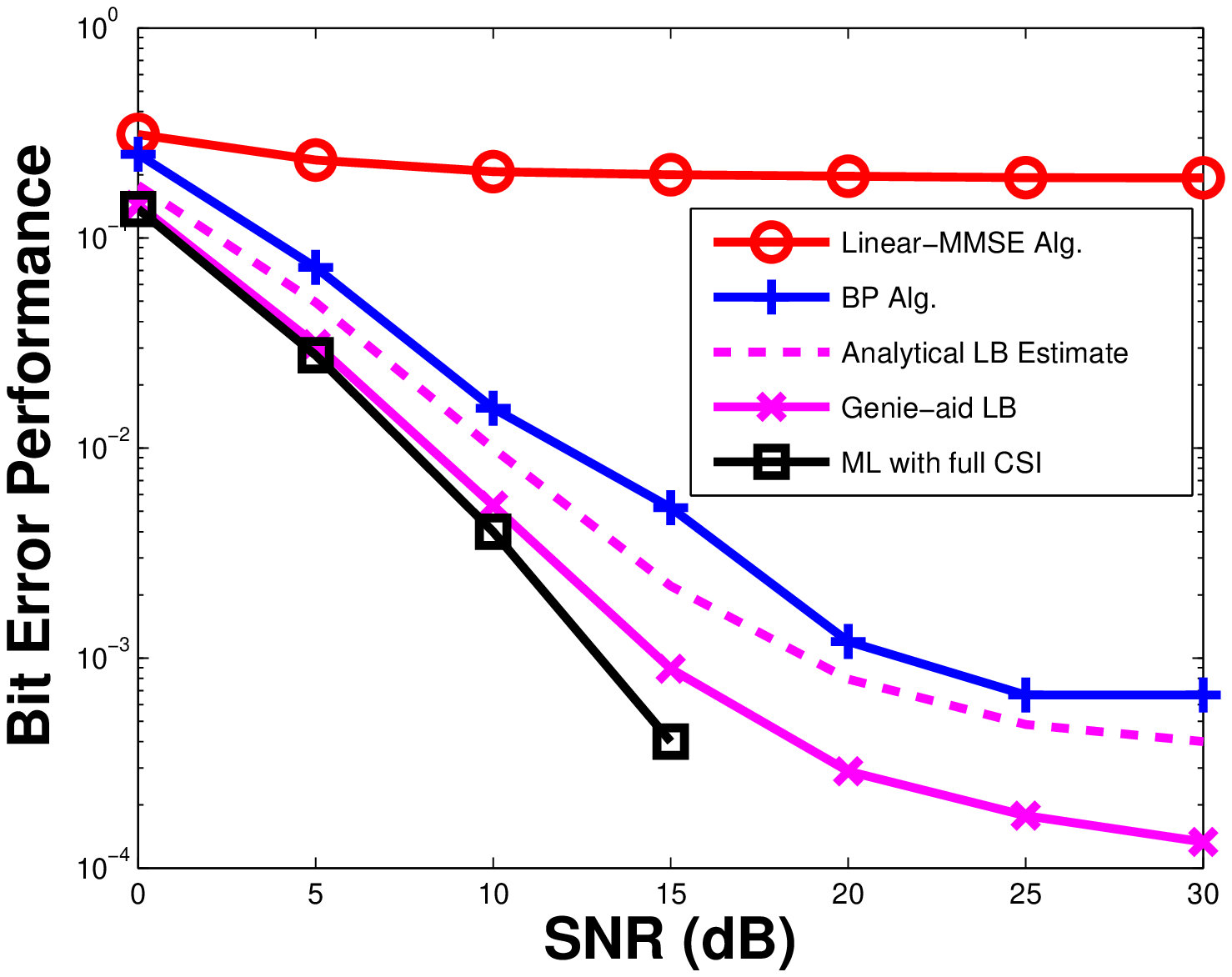}}
    \caption{The BER performance of the message-passing algorithm. The density
      of pilots is 25\%. (a) The power of the interference is 10 dB weaker than
      that of the desired user. (b) The power of the interference is 3 dB
      weaker than that of the desired user. (c) The power of the interference is identical to that of the
      desired user.}
    \label{fig:ber}
\end{figure*}


\section{Extensions and Complexity}\label{sec:ext}
\subsection{Extensions}\label{sec:fgraph:ext}
The message-passing approach applies to general multiple-input multiple-output
systems. For example, if $N_T$ transmit antennas are used by the desired user,
$N'_{T}$ transmit antennas are used by the interferer, and $N_R$ antennas are
used by the receiver, then the system can be described as
\begin{subequations}
  \begin{align}
    \vy_i &= \mH_i \vx_i + \mH'_i \vx'_i + \vn_i \label{eq:sys-g1}\\
    \mH_i &= \mF \mH_{i-1} + \mW_i \label{eq:ev-ga1} \\
    \mH'_i &= \mF' \mH'_{i-1} + \mW'_i \label{eq:ev-ga2}
  \end{align}
\end{subequations}
where $\vy_{i}(N_R\times 1)$, $\vx_i(N_T\times 1)$, $\vx'_i(N'_T\times 1)$ are
the received signal, desired user's signal and interfering signal, respectively,
at time $i$, the noise $\vn_i(N_R\times 1)$ consists of CSCG entries, and
$\mH_i(N_R\times N_T)$ and $\mH_i(N_R\times N'_T)$ are independent channel
matrices. Equations (\ref{eq:ev-ga1}) and (\ref{eq:ev-ga2}) represent the
evolution of the channels, where $\mF$ and $\mF'$ are in general square
matrices, and $\mW_i$ and $\mW'_i$ are independent CSCG noises.

Let $\vh_{j,i}$ represent the $j$-th column of $\mH_i$, and define
 \begin{align*}
    \vbg_i &= \left[\vh_{1,i}^\uT, \vh^\uT_{2,i}, \dots, \vh^\uT_{N_T,i},
      \vh'^\uT_{1,i}, \vh'^\uT_{2,i}, \dots, \vh'^\uT_{N'_T,i} \right]^\uT \\
    \vbu_i &= \left[\vw^\uT_{1,i}, \vw^\uT_{2,i}, \dots, \vw^\uT_{N_T,i},
      \vw'^\uT_{1,i}, \vw'^\uT_{2,i}, \dots,
      \vw'^\uT_{N'_T,i} \right]^\uT \\
    \mZ_i &=  \left[ \vx^\uT_i, \vx'^\uT_i \right]^\uT  \otimes \mI_{N_R} \\
    \mA &= \mathbb{E}\left[\vbg_{i} \vbg^\uH_{i-1} \right] \left( \mathbb{E}[\vbg_{i-1}
    \vbg^\uH_{i-1}] \right)^{-1} \\
    \mB &= \mathbb{E}[\vbg_{i} \vbu_i^\uH]\,.
  \end{align*}
Note that (\ref{eq:sys3}) and (\ref{eq:auto-reg3}) are still valid, where
$\alpha$ and $\sqrt{1-\alpha^2}\,\mQ$ are replaced by $\mA$ and $\mB$,
respectively. Therefore, with this replacement, the BP algorithm for this
general model remains the same.

We can also replace the Gauss-Markov model with higher order Markov models.  By
expanding the state space (denoted by $\mG_i$), we can still construct the
corresponding factor graph by replacing variable nodes $(\mH_i, \mH'_i)$ with
$\mG_i$, and a similar algorithm can be derived as before. Also, extensions to
systems with more than one interference can be similarly derived.

Furthermore, the proposed scheme can in principle be generalized to any signal
constellation and any space-time codes, including QPSK, 8-PSK, 16-QAM and
Alamouti codes. However, as the constellation size, the space-time codebook size or the
number of interferers increases, the complexity of the algorithm increases
rapidly, while the advantage over linear channel estimation vanishes because the
interference becomes more Gaussian due to central limit theorem. Thus the
algorithm proposed in this paper is particularly suitable for BPSK and QPSK
modulations, space-time codewords with short block length and a small number of
interferers. A detailed tally of the total complexity is given next.

\subsection{Complexity}\label{sec:fgraph:cmx}
With or without coding, the complexity of the message-passing receiver is linear
in the frame length, and polynomial in the number of antennas.

Suppose that there are $m$ channel coefficients, so $\vbg$ is a vector of
length $m$ (for the Gauss-Markov channel model $m=N_R(N_T+N'_T)$). The number of
receive antennas is $N_R$, the maximum number of Gaussian components we allow is
$C$, and the sizes of the alphabet of $x_i$ and $x'_i$ are $|\mathcal{A}|$ and
$|\mathcal{A'}|$, respectively. The complexity of computing
$p(\vbg_i|\vy^{i-1}_1)$ is then 
$O(C|\mathcal{A}||\mathcal{A'}| N^{a}_Rl)$,
where $l$ is the frame length and $N^{a}_R$ is due to the matrix inverse in
(\ref{eq:var}). The exponent $a$ depends on the 
particular inversion algorithm, and is typically
between two and three.\footnote{The value $a=2.37$ is established
in~\cite{Coppersmith87Matrix} for general matrices.
There has been recent progress on developing efficient algorithms for
  matrix computations~\cite{Cohn05Matrix} and the Hermite matrices
  in~\eqref{eq:var} may allow a further reduction in complexity.}
Similar complexity is needed to compute
$p(\vg_i|\vy^l_{i+1})$. To synthesize the results from the backward and forward
message passing via (\ref{eq:soft}), we need
$O(C^2|\mathcal{A}||\mathcal{A'}|m^{a}l)$ computations. Thus, the total
complexity for the uncoded system is
$O((CN^{a}_R+C^2m^{a})|\mathcal{A}||\mathcal{A'}|l)$.  To reduce the
complexity, one can reduce $C$, which causes performance loss.  One can also try
to approximate the matrix inverse (or equivalently, replace the Kalman filter
with a suboptimal filter).

For a coded system, the complexity of message-passing LDPC decoder is generally
linear in codeword length~\cite{Richardson01LDPC1}. With multiple frames coded
into one codeword, the decoder complexity is also linear in the frame
length. Suppose the number of EI exchanges between detector and decoder is
$I_{det}$. Then the overall complexity for the receiver is $O( (CN^{a}_R +
C^2m^{a})I_{det} |\mathcal{A}||\mathcal{A'}| l)$.


\section{Simulation Results}\label{sec:simu}
In this section, the model presented in Section~\ref{sec:model} with 
dual receive antennas~($N_R=2$) and BPSK signaling, is used for simulation. The
performance of the message-passing algorithm is plotted versus signal-to-noise
ratio $SNR=\sigma_h^2/\sigma_n^2$, where the covariance matrix of the noise is
$\sigma_n^2 \mI$. We set the channel correlation parameter\footnote{In
    Clark model, correlation between adjacent symbols is $.99$ corresponds to the
    scenario with the normalized maximum Doppler frequency approximately
    0.03. In the other words, $\alpha=.99$ corresponds to $300$~Hz of Doppler
    spread with symbol rate of $10$~Kbps.} $\alpha =.99$ and limit the maximum number of Gaussian
components to $8$. Within each block, there is one pilot in every $4$
symbols. For the uncoded system, we set the frame length to $l = 200$. For the
coded system, we use a $(500,250)$ irregular LDPC code and multiplex one LDPC
codeword into a single frame, \emph{i.e.}, we do not code across multiple frames.

\subsection{Performance of Uncoded System}
\subsubsection{BER Performance}
\begin{figure}[t]
  \centering
  \includegraphics[width=3.4in]{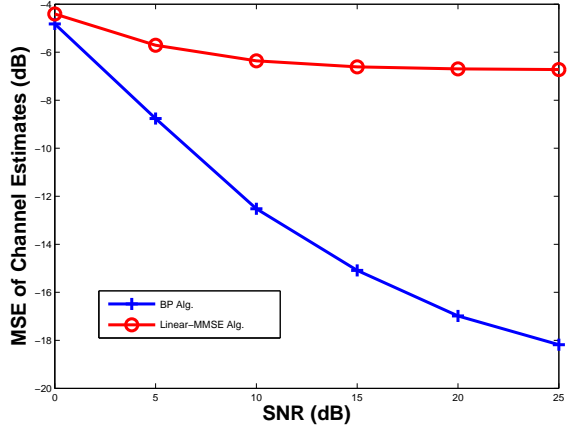}
  \caption{Channel estimation error with an interferer, which is 3 dB weaker than the desired signal. The density of pilots is 25\%.}
  \label{fig:ch_m}
\end{figure}

\begin{figure}
  \centering 
   \includegraphics[width=3.48in]{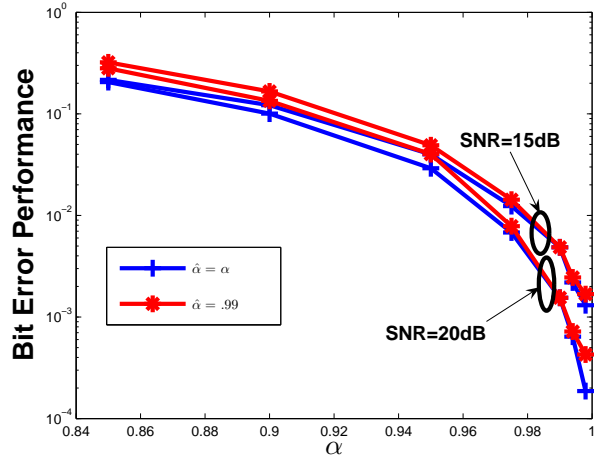}
  \caption{Performance
    v.s. different values of $\alpha$ with 3~dB weaker interference. For curves
    marked with ``+'', the receiver uses the true value of correlation coefficient
    \emph{i.e.}, $\hat{\alpha}=\alpha$. For curves marked with ``*'', receiver
    uses $\hat{\alpha}=.99$ regardless of the value of $\alpha$. }
   \label{fig:robust1}
\end{figure}

\begin{figure}
    \centering
    \includegraphics[width=3.48in]{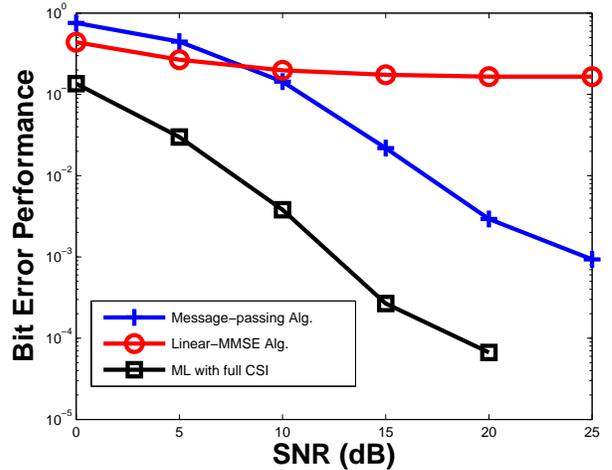}
    \caption{ Performance under the Clarke channel with normalized maximum
      Doppler frequency $0.02$ and 3~dB weaker interference.}
     \label{fig:robust2}
\end{figure}

Results for the message-passing algorithm with the Gaussian mixture messages
described in Section \ref{sec:fgraph} are shown in Figs.~\ref{fig:ber} to
\ref{fig:ldpc2}. We also show the performance of three other receivers for
comparison. The first is denoted by ``MMSE'', which estimates the desired
channel by taking a linear combination of adjacent received value. This MMSE
estimator treats the interference as white Gaussian noise. The second is the
genie-aided receiver described in Section~\ref{sec:ana}, denoted by ``Genie-aided
LB'', which gives a lower bound on the performance of the message-passing
algorithm. The third one is denoted by ``ML with full CSI'', which performs
maximum likelihood detection for each symbol assuming that the realization
of the fading processes is revealed to the detector by a genie, which lower
bounds the performance of all other receivers. We also plot the
approximation of the BER for the optimal genie-aided
receiver obtained from (\ref{eq:error}) using a dashed line.

Fig.~\ref{fig:ber} shows uncoded BER vs. SNR, where the power of the
interference is $10$ dB weaker, $3$ dB weaker and equal to that of the desired
user, respectively. The message-passing algorithm generally gives a significant
performance gain over the MMSE channel estimator, especially in the high SNR
region. Note that thermal noise dominates when the interference is
weak. Therefore, relatively little performance gain over the MMSE algorithm is
observed in Fig.~\ref{fig:ber_w}. In the very low SNR region, the MMSE
algorithm slightly outperforms the message-passing algorithm, which is probably
due to the limitation on number of Gaussian components.

The trend of the numerical results shows that the message-passing algorithm
effectively mitigates or partially cancels the interference at all SNRs of
interest, as opposed to suppressing it by linear filtering. We see that there is
still a gap between the performance of the message-passing algorithm and that of
the genie-aided receiver. The reason is that revealing the channel coefficients
enables the receiver to detect the symbol of the interferer with improved
accuracy. Another observation is that the analytical estimate is closer to the
message-passing algorithm performance with stronger interference.

\subsubsection{Channel Estimation Performance}
The channel estimate from the message-passing algorithm is much more accurate
than that from the conventional linear channel estimation. 
Fig.~\ref{fig:ch_m} shows the mean squared error for the channel estimation
versus SNR where the interference signal is 3 dB weaker than the
desired signal and one pilot is used after every three data symbols.  
Note that the performance of the linear estimator hardly improves as
the SNR increases because the signal-to-interference-and-noise ratio is no
better than 3~dB regardless of the SNR.  This is the underlying reason for the
poor performance of the linear receiver shown in Fig.~\ref{fig:ber}.

\subsection{The Impact of Imperfect Knowledge of Channel Statistics}
Although the statistical model for the channel is usually determined \emph{a
  priori}, the parameters of the model are often based on on-line estimates,
which may be inaccurate. The following simulations evaluate the robustness of
the receiver when some parameters, or the model itself is not accurate. The
simulation conditions here are the same as for the previous uncoded system with
3~dB weaker interference. Fig.~\ref{fig:robust1} plots the BER performance
against the correlation coefficient $\alpha$, while the receiver uses
$\hat{\alpha}$ instead. It is clear that the mismatch
in $\alpha$ causes little degradation. The result
of a similar experiment is plotted in Fig.~\ref{fig:robust2}, where the receiver
assumes the Gauss-Markov model, while the actual channels follow the Clarke
model~\cite[Ch. 2]{Tse05book}.\footnote{We set $\hat{\alpha}$ according to the
  auto-correlation function for the Clark model.} We see that the
message-passing algorithm still works well. In fact, as long as the channel
varies relatively slowly, modeling it as a Gauss-Markov process leads to good
performance.

\subsection{Coded system and the Impact of Message-passing
Schedule}
Consider coded transmission using a $(500,250)$ irregular LDPC code\footnote{The
  left degree parameters are $\lambda_3 = .9867$, $\lambda_4=.0133$; the right
  degree parameters are $\rho_4=.0027$, $\rho_5=.0565$, $\rho_6=.8332$,
  $\rho_7=.1023$, $\rho_8=.0053$. For the meaning of the parameters, please
  refer to~\cite{Richardson01LDPC2}.}  and with one LDPC codeword in each frame,
\emph{i.e.}, no coding across multiple frames. Since we insert one pilot after
every $3$ symbols, the total frame length is $667$ symbols. For the message-passing algorithm, let $I_{det}$ denote the total number of EI exchanges between decoder and
detector, and $I_{dec}$ denote the number of iterations of the LDPC decoder
during each EI exchange. Different values for pair $(I_{det}, I_{dec})$
correspond to different message-passing schedules. 

\begin{figure}
  \centering
  \includegraphics[width=3.5in]{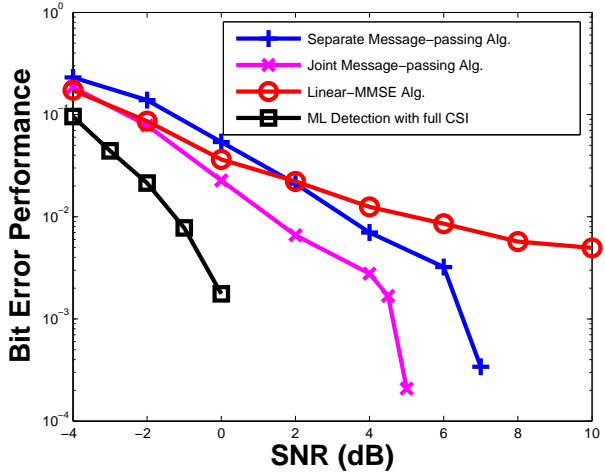}
  \caption{The BER performance for the system with a $(500, 250)$ irregular LDPC
    code. The interference is 3 dB weaker than the desired signal. The density
    of pilots is 25\%. ``ML Detection with full CSI'' refers to ML detection
    with full CSI followed by BP-based LDPC decoding, which serves as a
    performance benchmark. } 
  \label{fig:ldpc1}
\end{figure}

In Fig.~\ref{fig:ldpc1}, we present the performance of two
message-passing schedules: (a) $I_{det}=1$ and $I_{dec}=50$ denoted by
``Separate Message-passing Alg.'', \emph{i.e.}, the receiver detects the symbol
first, then passes the likelihood ratio to the LDPC decoder without any further
EI exchanges (separate detection and decoding), and (b) $I_{det}=5$ and
$I_{dec}=10$, denoted by ``Joint Message-passing Alg.'', \emph{i.e.}, there are
five EI exchanges and the LDPC decoder iterates $10$ rounds in between each EI
exchange. For the other two receiver algorithms, the total number of iterations
of LDPC decoder are both~$50$. As shown in Fig.~\ref{fig:ldpc1}, the
message-passing algorithm preserves a significant advantage over the traditional
linear MMSE algorithm and the joint message-passing algorithm gains even more.

\begin{figure}
  \centering
      \includegraphics[width=3.5in]{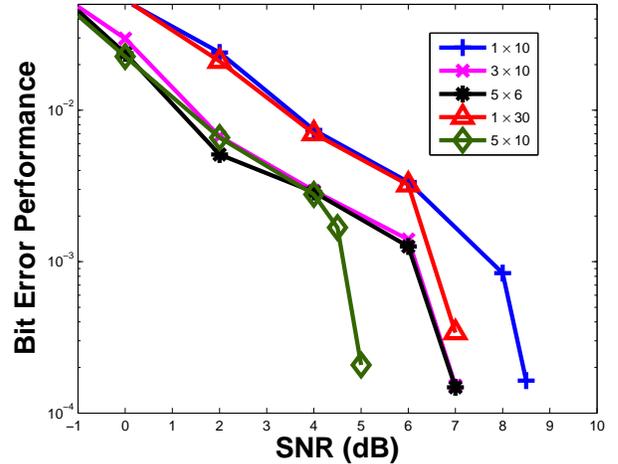}
    \caption{The impact of different message-passing schedules.}
 \label{fig:ldpc2}
\end{figure}

The performance with different message-passing schedules is shown in
Fig.~\ref{fig:ldpc2}. where we fix total number of LDPC iterations
$I_{det}\times I_{dec}$.  Generally speaking, if $I_{dec}$ or $I_{det} \times
I_{dec}$ is fixed, more EI exchanges lead to better performance. We also observe
that when $I_{dec}$ is relatively large, say $30$, the performance gain from EI
exchanges is small. The reason is that when $I_{dec}$ is large, the output of
the LDPC decoder ``hardens'', \emph{i.e.}, the decoder essentially decides what
each information bit is. When the EI is passed to the detector, all symbols look
like pilots from the point of view of the detector. Therefore, there is not much
gain in this case.
 
\subsection{The Impact of Mixture Gaussian Approximation}  
As previously mentioned, the number of Gaussian components in the messages
related to the fading coefficients grows exponentially. For implementation, we
often have to truncate or approximate the mixture Gaussian message. In this
paper, we keep only a fixed number of components with the maximum
amplitudes. The maximum number of components clearly has some impact on the
performance. Here we present some numerical experiments to illustrate this
effect.

\begin{figure}
  \centering
  \includegraphics[width=3.5in]{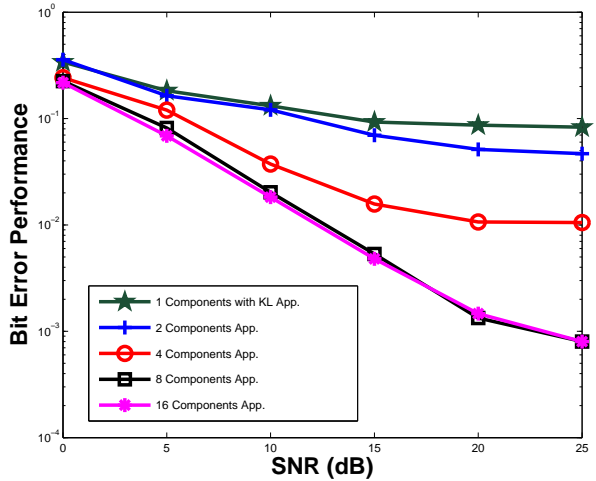}
  \caption{The BER performance with different number of components in the messages. The interference
    is assumed to be 3~dB weaker than the desired signal. The  density of pilots is $25\%$.}
  \label{fig:maxc_b}
\end{figure}

When the pilot density is high, say $50\%$, there is no need to keep many Gaussian
components in each message. In fact, keeping two components is essentially
enough. However, when the pilot density is lower, say $25\%$, the situation is
different. Fig.~\ref{fig:maxc_b} shows the BER performance when we keep
different numbers of Gaussian components in the message-passing algorithm where
the pilot density is $25\%$. For this case, we need $8$ components for each
message passing step. Indeed, the lower the pilot density, the more Gaussian
components we need to achieve the same performance. When the pilot density is
low, we must keep a sufficient number of components, corresponding to a
sufficient resolution for the message. Roughly speaking, the number of Gaussian
components needed is closely related to the number of hypotheses arising from
symbols between the symbol of interest and the nearest pilot.

For a single-user system, previous studies indicate that a single Gaussian
approximation of the messages is sufficient, \emph{e.g.},
\cite{Niu05BP}. Fig.~\ref{fig:maxc_b} shows that this is not the case
for the system considered with one dominant interferer. 
For the simulation with the single Gaussian approximation
all Gaussian components are combined into one at each message-passing stage 
according to the minimum divergence criterion~\cite{Kurkoski08BP}. 
As expected, the performance with the single Gaussian approximation
is relatively poor.
Namely, consider the posterior probability, or conditional PDF, of
$(\vh_i, \vh_i')$, which is the message passed along the graph. 
Due to the lack of
pilots for estimating $\vh'_i$, there is inherent ambiguity of its polarity so
that the posterior of $\vh'_i$ is always symmetric around the origin even with
an exact message-passing algorithm. Consequently, any approximation with 
a single Gaussian function can do no better than 
treating $\vh'_i$ as a zero-mean Gaussian random vector.
This is equivalent to treating $\vh'_ix'_i$ as Gaussian noise, 
which leads to poor performance.

\section{Conclusion}\label{sec:con}
A novel architecture based on graphical models and belief propagation has been
proposed for joint channel estimation, interference mitigation and
decoding. Such joint processing is facilitated by efficient iterative
message-passing algorithms, where the total complexity is essentially the sum of
the complexity of the components, rather than their product as is typical in
joint maximum likelihood receivers. In the presence of time-varying Rayleigh
fading and a strong co-channel interference, the message-passing algorithm
provides a much lower uncoded error floor than linear channel estimation. The
results with LDPC codes show at least 5~dB gain for achieving acceptable
bit-error rates. Also, this gain is robust with respect to mismatch in channel
statistics.

We have considered only two users with multiple receive antennas. Although this
is an important case, and the approach can be generalized, there may be
implementation (complexity) issues with extending the algorithm. For example, if
we have more than one interferer or use larger constellations, the number of
hypotheses at each message-passing step increases significantly. To maintain a
target performance, we need to increase the number of Gaussian components in
each step accordingly. Therefore, the complexity may significantly increase with
these extensions. Finally, the algorithm is difficult to analyze. While our
results give some basic insights into performance, relative gains are difficult
to predict.

Directions for future work include extensions to MIMO channels (where channel
modeling within the message-passing framework becomes a challenge) as well as
implementation issues including methods for reducing complexity.  Extending the
message-passing approach to equalization of frequency selective channels with
interference is also an interesting direction. For example, the narrowband
model~\eqref{eq:sys} considered here could be viewed as an OFDM system with a
number of sub-channels. (The receiver algorithm should then be modified to
account for correlations across sub-channels.) Alternatively, message-passing
approach could be combined with adaptive equalization in the time domain.

\section*{Acknowledgment}
We thank Dr. Kenneth Stewart for helpful discussions during the course of this project, and Mingguang Xu for discussions and help with the simulations.  We would also like to thank the anonymous reviewers for useful comments which helped to improve the presentation of the paper.

\bibliographystyle{IEEEtran} \bibliography{IEEEabrv,zybibset,zybibset_tmp}
\vfill

\end{document}